# LWRP: Low Power Consumption Weighting Replacement Policy using Buffer Memory

Mr. S.R.Bhalgama[1], Mr.C.C.Kavar[2], Mr. S.S.Parmar[3]

[1]PG Student, C. U. Shah College of Engineering and Technology, Gujarat, India
[2]Asst. Professor, C. U. Shah College Of Engineering & Technology, Gujarat, India.
[3]Asst. Professor, C. U. Shah College Of Engineering.& Technology, Gujarat, India.

**ABSTRACT:** As the performance gap between memory and processors has increased, then it leads to the poor performance. Efficient virtual memory can overcome this problem. And the efficiency of virtual memory depends on the replacement policy used for cache. In this paper, our algorithm not only based on the time to last access and frequency index but, we also consider the power consumption. We show that Low Power Consumption Weighting Replacement Policy (LWRP) has better performance and low power consumption.

***Keywords*** *– LFU, LRU, Memory management, replacement policy, weighting replacement policy.*

## I. INTRODUCTION

Caching is a well known technique towards optimize performance that is widely used in computer systems. Increasing cache size gives better results but, it is very expansive. Thus we must have to find out another alternative to improving cache. And replacement policy is one of them [1, 3]. Cache performance could be changed by implementing different replacement policy. Time access for cache is 10 times smaller than main memory [3]. Replacement policy specifies the parameters and based on these parameters a new page will be replaced with existing one. Most replacement algorithms in use are based on the time to last access and frequency index but, it fails in some applications [1, 2, 3]. Other new replacement policy perform better way but, most of them hard to implement.

Three common caches have been used: direct mapped in which each block from the main memory is mapped to unique cache block. This organization of cache doesn't require a replacement policy; fully associative in which each block from the main memory can be mapped to any of the cache block; set associative in which the cache is split into many sets. Any block from the main memory is mapped only to the block of a certain set. Direct method is nothing but, one-way set associative [2, 3].

The basic idea of our algorithm is to weight and rank pages based on the three parameters: time to last access, frequency and the power consumption. So, pages that are more recent, more used and low power consumption is rank higher. Therefore the probability of replacing pages with smaller weight is more than the one with higher weight [3].

The remainder of this paper is organized as follow: In the next section, we review the existing cache replacement policy. In section III, we describe the energy model. In section VI, we describe the LWRP cache replacement algorithm. Its simulation and result discussed in section V. In section VI we include conclusion of our work.

## II. RELATED WORK

Caching is a technique that widely applied in many computer science applications. Database and operating system are two most important ones. Now days, World Wide Web is becoming another popular area of caching [1, 2, 3]. The objective of this paper is reducing the power consumption by reducing cache misses. So we start with related work for cache replacement algorithm then introduce related work for power consumption [2].

The most popular cache replacement algorithms are first in first out (FIFO), most recently used (MRU), least recently used (LRU), and least frequently used (LFU). FIFO algorithm replace block that was first referenced. The MRU algorithm replaces the block that referenced most. The LRU algorithm replaces the block that has not been used for the long time. The LFU algorithm replaces the block that was least frequently referenced [2,3,4].

Many different algorithms have been proposed by researcher to improve the performance of cache memory. Some of this algorithms are the frequency based replacement (FBR), second chance frequency-least recently used (SF-LRU), LRU-K algorithm, least recently frequently used (LFRU). LRU-K algorithm evict block based on the time of the Kth –to-last reference to the block [5]. The FBR algorithm is a hybrid combination of LRU and LFU. The LRFU replacement algorithm based on recency and frequency (CRF) value and replaces the block with minimum CRF value [6].

There are drawbacks of the above mentioned algorithms. LRU is uses only the time of the most recent reference to each block and cannot determine the frequency of the block. The LFU on other hand cannot determine the recency of the block. The LRU-K considers only the Kth reference. All other algorithms such as LRFU, FBR has a lot of implement overhead [3, 7].

Now our main focus is power consumption. Lot of work have been done on reducing the power consumption of cache memory. Memik et al. proposed a victim cache structure to reduce the number of accesses to more power consuming structures [8]. Nicolaescu et al. propose a technique utilizing cache line address locality to determine the cache way prior to the cache access [9]. So our algorithm reduced the power consumption using the special buffer.

### III. ENERGY MODEL

Several energy models have been proposed for caches. The energy model developed in [11] is our base. Where energy is given by:

$$Energy = (HR * EnergyHit) + (MR * EnergyMiss) \quad (1)$$
$$EnergyHit = EnergyDecoder + EnergyCellArray \quad (2)$$
$$EnergyMiss = EnergyHit + EnergyAccessMemory \quad (3)$$

Where HR= Hit Rate and MR=Miss Rate. The *EnergyCellArray* is the energy in the cell arrays, *EnergyAccessMemory* is the energy require to access data in main memory. And *EnergyDecoder* is the energy in the decoder. Energy require to access the data from main memory consume the majority energy of the overall power cost. So that *EnergyMiss >> EnergyHit*, and it is clear that if miss rate reduction is achieved, then energy consumption is reduced. More information on the complete model can be found in [11].

Therefore this prove that our approach to the power consumption problem is correct and when we achieve a better hit rate and reduce the number of misses we are actually reducing the power consumption of the system.

### IV. LWRP ALGORITHM

In this section, we introduce a cache replacement algorithm based on the energy model for reduce the power consumption

1. Power consumption minimization based on energy model

$$T(Hit) = T(EnergyHit) \quad (4)$$
$$T(Miss) = T(EnergyMiss) \quad (5)$$
$$= T(EnergyHit) + T(EnergyAccessMemory)$$

Where T= time.

Therefore *T(Miss)>> T(Hit)*. If we minimize the *T(EnergyAccessMemory)* then we can reduce the *T(Miss)*. For that we propose a new strategy for reduce *T(EnergyAaccessMemory)*.

Fetching the data from disk require at least 1000 more power consumption than buffer [10]. We used a special buffer in this algorithm to reduce access time. If cache is full for any miss operation, then we evict block from cache and move it to buffer. So all the evicted block from cache moved to the buffer until the buffer is full shown in Fig1

However, if there is a miss in cache but, hit in buffer, also known as Partial miss, it require less time than actual miss.

$$T(PartialMiss) = T(EnergyHit) + T(EnergyAccessBuffer) \quad (6)$$

Where T=time and T(EnergyAccessBuffer) << T(EnergyAccessMemory).

Therefore by reducing the time of miss we can minimize the power consumption. Fig. 2 shows the pseudo code for this strategy.

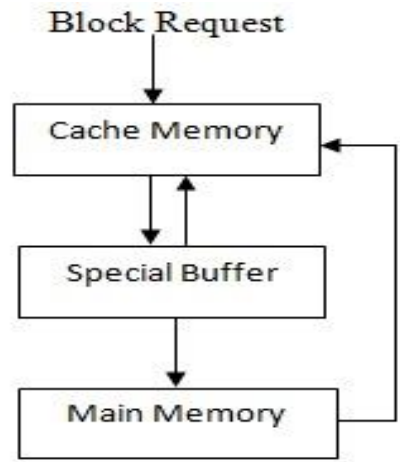

Fig. 1 Cache replacement strategy using buffer

```
IF( Miss in cache )
    IF( Miss in buffer )
        // move block from main memory to cache
        IF( free frame )
            Move block;
        ELSE
            Execute LWRP;
            Move block;
    ELSE
        //move block from buffer to cache
        IF( free frame )
            Move block;
        ELSE
            Execute LWRP;
            Move block;
```

Fig. 1  Pseudo code of LWRP algorithm

2. Replacement strategy

We assume the size of the blocks is equal and the replacement algorithm is used to manage finite number of blocks [1,3]. A hit occur when there is a reference to a block in the cache. A partial hit occur when there is a reference to a block in the special buffer. When we have a reference to a block not in cache and buffer, a miss will occur. When a miss occur and no free frame available in cache, we must replace a new block to evicted block.

For replacement we consider three factors. Let $R_i$ be the recency counter of block $i$ and *HitEnergy* be the energy of hit of block. And consider the time difference $\Delta T_i = T c_i - T p_i$ where $T c_i$ is the time of last access and $T p_i$ is the time of penultimate access. Then the weighting value of block can be calculated as [1]

$$W_i = \frac{R_i}{(HitEnergy)_i * \Delta T_i} \quad (7)$$

When a new block $k$ is placed in the cache then all the above mentioned parameters in weighting function must be set, and followed by setting $R_k$ to 0, *Hit_Energy* to 1 and $\Delta T_k$ to 1. We assume that initial value of *Hit_Energy* and $\Delta T_k$ to 1 because the time between each cache reference to a block would be at least 1 in its minimum case. Based on these three parameters we calculate the weighting value of every block.

In every access of cache, if reference block $j$ is in the cache then a hit is occurred and our policy will work as follow:

1) $R_i$ will be changed to $R_i + 1$ for every $i \neq j$.
2) For $i = j$ first we assign,
   $\Delta T_i = R_i$,
   *HitEnergy* = *HitEnergy*(*old*) + *HitEnergy*(*current*)

And then $R_i=0$.

But if reference block $j$ is in the buffer then partial hit is occurred and our policy work as follow:

1) Moves block j from buffer to cache.
2) Set all weighting parameters of block $j$ to their initial values.

If reference block $j$ is not in the cache and not in the buffer and no free frame is available then miss occurs. Then our policy work as follow:

1) Choose the block $k$ which has a highest weighting value.
2) Change $R_i$ to $R_i + 1$ for every $i \neq k$.
3) Replace a new reference block with block $k$.
4) Set all weighting parameters of block $k$ to their initial values.

The weighting function of block will update in every access to cache. We are only describing how our algorithm works, but we have not discussed how to implement in system mechanism. One important factor in replacement algorithm is its overhead in the system [1, 3].

LWRP require storing three parameters to work and it will add space overhead: first algorithm require space for counter $Ri$, need a space for a *HitEnergy* and third it needs a space for counter $\Delta Ti$. As well as space for weighting value $Wi$. Calculating weighting value after every access to cache will increase the time overhead to system.

## V. CONCLUSION

In this paper we proposed a low power consumption weighting replacement policy in which we can minimize the energy of replacement algorithm.

In future we can simulate our policy and compare it with other policies like LRU and LFU. The simulator is a program was designed to run traces and implement different replacement algorithm. The obtained hit ratio depends on the replacement algorithm, locality of reference and the cache size.